\documentclass{article}
\usepackage{arxiv}
\usepackage[utf8]{inputenc} % allow utf-8 input
\usepackage[T1]{fontenc}    % use 8-bit T1 fonts
\usepackage{hyperref}       % hyperlinks
\usepackage{url}            % simple URL typesetting
\usepackage{microtype}      % microtypography
\usepackage{graphicx}

\title{Optical Properties of LiNbO\textsubscript{2}
Thin
Films}

\author{ T. Kurachi \\
	Far-IR R\&D Center\\
	University of Fukui\\
	Fukui 910-8507, Japan\\
	\And
T. Yamaguchi\\
	Far-IR R\&D Center\\
	University of Fukui\\
	Fukui 910-8507, Japan\\
\And
E. Kobayashi\\
Advanced Interdisciplinary Science \& Technology\\
	University of Fukui\\
	Fukui 910-8507, Japan\\
\And
\href{https://orcid.org/0000-0001-8685-9606}{\includegraphics[scale=0.06]{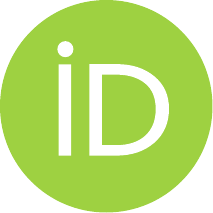}\hspace{1mm}T. Soma}\\
Department of CS\&E\\
Tokyo Institute of Technology\\
Tokyo 152-8552, Japan\\
\And
\href{https://orcid.org/0000-0003-0300-4712}{\includegraphics[scale=0.06]{./img/orcid.pdf}\hspace{1mm}A. Ohtomo} \\
Department of CS\&E\\
Tokyo Institute of Technology\\
Tokyo 152-8552, Japan\\
\And
\href{https://orcid.org/0000-0002-5706-8909}{\includegraphics[scale=0.06]{./img/orcid.pdf}\hspace{1mm}T. Makino} \\
	Far-IR R\&D Center\\
University of Fukui\\
Fukui 910-8507, Japan\\
	\texttt{tmakino@u-fukui.ac.jp} \\
}

\renewcommand{\shorttitle}{Optical Properties of LiNbO\textsubscript{2}}

\hypersetup{
pdftitle={Optical Properties of LiNbO\textsubscript{2}},
pdfsubject={cond-mat},
pdfauthor={T. Kurachi, T.~Makino},
pdfkeywords={Physica B 621 413259 (2021)},
}

\begin{document}
\maketitle

\begin{abstract}
The complex dielectric functions of LiNbO\textsubscript{2} were
determined using optical transmittance and reflectance spectroscopies at
room temperature. The measured dielectric function spectra reveal
distinct structures at several bandgap energies. The bandgaps (exciton
resonances) in the spectrum were observed at \emph{ca.} 2.3, 3.2, 3.9,
and 5.1 eV, respectively. These experimental data have been fit using a
model dielectric function based on the electronic energy-band structure
near critical points plus excitonic effects. The features of measured
dielectric functions are, to some extent, reproduced quantitatively by
an \emph{ab-initio} calculation including the interaction effects
between electrons and holes.
\end{abstract}

\section{1. Introduction}\label{introduction}

Lithium niobate is a member of the transition-metal oxide compounds. At
normal temperature and pressure, it crystallizes in a form of
three-dimensional LiNbO\textsubscript{3}. Lithium niobate can also
crystallize in the layered LiNbO\textsubscript{2} (LNO) modification,
although it is metastable under normal conditions {[}1-2{]}. By using
the epitaxial growth techniques such as pulsed laser deposition, it is
now possible to grow LiNbO\textsubscript{2} thins films without the need
for rather restricted conditions such as a reducing atmosphere
{[}3-6{]}. This compound is known to be several intriguing properties
such as the superconductivity {[}7-10{]} in the delithiated phase at
approximately 5 K and the high ionic conductivity suitable for potential
applications as a battery material {[}11-13{]}.

Although stoichiometric LNO is a ``mother material'' of the delithiated
counterpart having compatibility of transparency and the
superconductivity, the little is known about the optical response of
this compound. LNO is a direct gap semiconductor {[}14-17{]}. The
bandgap energies were evaluated to be \emph{ca.} 2.0 eV both
experimentally {[}18{]} and theoretically {[}14{]}. On the other hand,
the nature of these optical transitions has not been clarified yet for
this material. This is partly because the experimental information is
based on a diffused reflectance spectrum taken on a specimen in powdered
form {[}18{]}. In this case, the conversion to the dielectric functions
is very difficult. Generally, to draw electronic-structure-related
knowledge from measured data, analysis based on the standard critical
point (SCP) modeling {[}19{]} has been adopted so far. According to the
SCP theory, the lineshape is considered to reflect that of the spectral
distribution in the joint density of states of the valence and
conduction bands. The lineshape of the joint density of states strongly
depends on the dimensionality and the Lynch index which corresponds to
the number of negative components in the effective mass vectors. Gaining
insight into the nature of the optical transitions is attributed to the
determination of the parameters such as the dimensionality and the Lynch
index. Such a parameterization has not been, however, done so far for
this material. These knowledges are very important for \emph{e.g.}
device applications.

In this work, we study the optical properties of LNO thin films. We
present dielectric function spectra deduced from transmission and
reflectance spectra at room temperature (RT) between 1.5 and 6.5 eV. A
method is also described for calculating the spectroscopic distribution
of the dielectric function of LNO where the relevant models are
connected with the electronic energy-band structures of the compound.
For comparison, we also show \emph{ab-initio} calculation results with
the \emph{GW} level under linear-response approximations.

\section{2. Theoretical Model}\label{theoretical-model}

\textbf{2.1 Electronic energy-band structure of LiNbO\textsubscript{2}}

Several groups have reported the calculated results on the
electronic-energy band structures in LNO {[}14-17{]}. We also executed
local-density approximation (LDA) calculations by ourselves. The
reproducible results are obtained as shown in Fig. 1. To ensure
computing cost-effectiveness for regression analysis, we attempted to
reproduce the measured dielectric functions with critical points (CP's),
the number of which is as small as possible. From the
model-dielectric-function (MDF)-based regression analysis for the
measured data, the energies of the CP's were evaluated to be \emph{ca.}
2.3, 3.2, 3.9, and 5.1 eV. These transition energies are shown in Fig. 1
with vertical arrows and are labeled as \emph{E}\textsubscript{0},
\emph{E}\textsubscript{1}, \emph{E}\textsubscript{2}, and
\emph{E}\textsubscript{3}. It should be noted that, for the assignments
of the vertical arrows in Fig. 1, we looked for the corresponding direct
transitions lower in energy if we could not find a transition having the
same energy with the experiment. In the analysis, we neglect the
following aspects for simplicity: (1) transitions above the
\emph{E}\textsubscript{3} gap, (2) the contributions from indirect
transitions due to their extremely weaker nature in their intensities,
and (3) the spin-orbit effects and the related splitting of the bands
(conventionally denoted as $\Delta$ in the literature).

\begin{figure}
\includegraphics[width=0.8\linewidth]{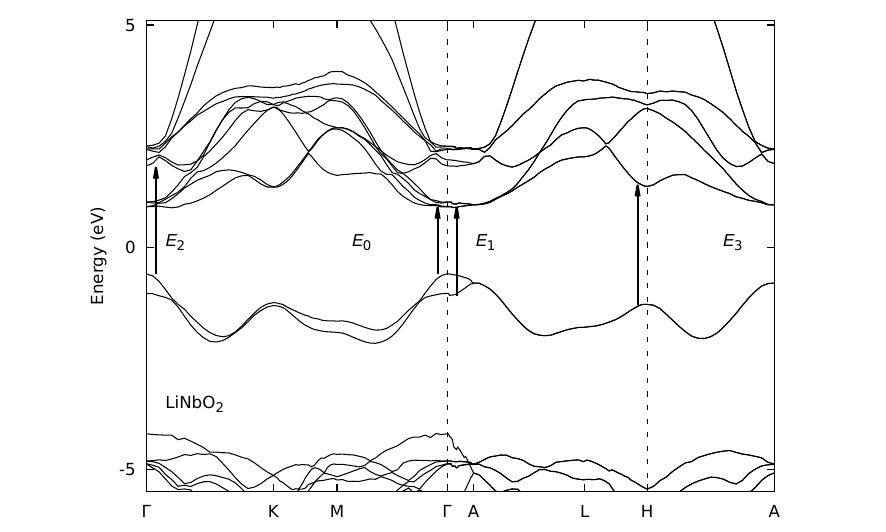}
\caption{Electronic energy-band structure of LiNbO\textsubscript{2}. }
\end{figure}

\textbf{2.2 Model dielectric functions}

We performed the line-shape analysis using the MDF approach {[}20{]}.
Here, we summarize dielectric functions as a form of a complex function.
To obtain $\epsilon$\textsubscript{1}($\omega$), one can take its real parts, while for
$\epsilon$\textsubscript{2}($\omega$), imaginary parts should be taken.

The \emph{E}\textsubscript{0}, \emph{E}\textsubscript{1},
\emph{E}\textsubscript{2} CP's may be of the three-dimensional (3D)
\emph{M}\textsubscript{1} type. The index \emph{i} in the
\emph{M}\textsubscript{i} notation corresponds to earlier-mentioned
Lynch index. As shown in Fig. 2, the overall feature of the
$\epsilon$\textsubscript{1}($\omega$) spectrum is characterized with monotonically
decreasing behavior concerning the photon energy, accompanied with
optical anomalies at the CP's. To reproduce this behavior, the
assumption of \emph{M}\textsubscript{1} type is more appropriate.
Because the \emph{M}\textsubscript{1} CP's longitudinal effective mass
is much larger than its transverse counterpart, one can treat these 3D
\emph{M}\textsubscript{1} CP's as a two-dimensional (2D) minimum
\emph{M}\textsubscript{0}. It is known that the equation in the
approximation of the 2D \emph{M}\textsubscript{0} CP gives a series of
Wannier-type excitons if taking the excitonic effects into account
{[}21{]}. In other words, one can neglect the one-electron contribution
to the dielectric function in the case of the 3D
\emph{M}\textsubscript{1} CP. Due to its layered structure, LNO is
expected to be a material where the electron-hole correlations are
rather strong. Rough calculation on the binding energy of the
Wannier-type exciton yielded in \emph{ca.} 70 meV for LNO by using the
reported reduced effective mass (0.57 \emph{m}\textsubscript{0}) and
background dielectric constant of \emph{ca.} 10.3 {[}14{]}. The notation
\emph{m}\textsubscript{0} means electron's mass in rest. Besides, the
excitonic effect could not be neglected in the case of 3D
\emph{M}\textsubscript{1} (2D \emph{M}\textsubscript{0}) CP for many
elemental and compound semiconductors {[}20,22-25{]}. Thus, we can
justifiably take this effect into account for the analysis. For
\emph{e.g.} the \emph{E}\textsubscript{0} feature, the contribution of
these excitons to $\epsilon$($\omega$) can be now written with Lorentzian lineshape as:

\begin{equation}
\epsilon(\hbar \omega) = \sum_{n = 1}^{\infty}\frac{1}{(2n - 1)^{3}}\left( \frac{A_{0}}{- i\Gamma_{1} + E_{0} - G_{0}/(2n - 1)^{2} - \hbar \omega} \right)
\end{equation}

where \emph{A}\textsubscript{0} is a constant corresponding to the
strength parameter, \emph{E}\textsubscript{0} is the energy of the CP,
\emph{G}\textsubscript{0} is the 2D binding energy of exciton. The value
of $\Gamma$\textsubscript{0} is the broadening parameter. Because the
ground-state exciton term occupies almost 95\% of the total oscillator
strength, we neglected the excited-state terms ($n \ge 2$) in the
current analysis.

The \emph{E}\textsubscript{3} peak is difficult to analyze as it does
not correspond to a single, well defined CP. Thus, the
\emph{E}\textsubscript{3} structure has been characterized as a damped
harmonic oscillator:

\begin{equation}
\epsilon(\hbar \omega) = \frac{A_{3}}{1 - \chi_{3}^{2} - i\Gamma_{3}\chi_{3}}
\end{equation}

with $\chi$\textsubscript{3} = \emph{E}/\emph{E}\textsubscript{3}, where
\emph{A}\textsubscript{3} is the strength parameter and
$\Gamma$\textsubscript{3} is a dimensionless broadening parameter.

\section{3. Experimental and Calculation
Procedures}\label{experimental-and-calculation-procedures}

LNO thin films were grown on MgAl\textsubscript{2}O\textsubscript{4}
(111) substrates using pulsed laser deposition (PLD) method with KrF
excimer laser. The growth was conducted in vacuum at RT. After
deposition, the films were annealed \emph{in situ} at substrate
temperature of 800 \textsuperscript{o}C under a chamber pressure of 0.1
mtorr set by continuous flow of Ar/H\textsubscript{2} gas. Films were
capped by several-nanometer-thick alumina films using PLD at RT in
vacuum for avoiding reactions with air. The description of detailed
growth process can be found elsewhere {[}6{]}. The film thickness is
approximately 150 nm.

Optical transmittance and reflectance were measured with an
ultraviolet-visible spectrometer at RT. Then, we converted these
spectral intensities to those of the complex dielectric functions
{[}26{]}. Here, the refractive index of the substrate is assumed to be
independent of the photon energy because the bandgap of
MgAl\textsubscript{2}O\textsubscript{4} is significantly wider than the
maximum energy of the measurement range (i.e., 6.5 eV).

Structural information of LNO has been reported by several experimental
groups, leading to the assignment of space group
\emph{P6\textsubscript{3}/mmc} (No.~194). We performed \emph{ab-initio}
calculation using the plane-wave basis set PWscf package of Quantum
ESPRESSO {[}27-28{]} to evaluate the electronic-energy-band structures.
For the lattice constant of LiNbO\textsubscript{2}, \emph{a}=2.938 \AA and
\emph{c}=10.596 \AA were used. The 3D Brillouin zone was integrated using
a $7 \times 7 \times 2$ k-point grid. We have set the energy cutoff for the
plane-wave basis to 30 Ry, and have used the Perdew--Burke--Ernzerhof
(PBE) functional, which belongs to the class of generalized gradient
approximation (GGA) functionals. In previous works, similar calculations
have been made {[}14-15{]}, and qualitatively coincided results could be
obtained by ourselves. For the optical spectra calculations such as
dielectric functions, we used Respack \emph{ab-initio} package
{[}29-30{]}. This package solves Bethe-Salpeter-like equations
numerically under single-excitation configuration-interaction (SECI)
treatment. By doing that, the interactions between electrons and holes
are taken into account.

\begin{figure}
\includegraphics[width=0.7\textwidth]{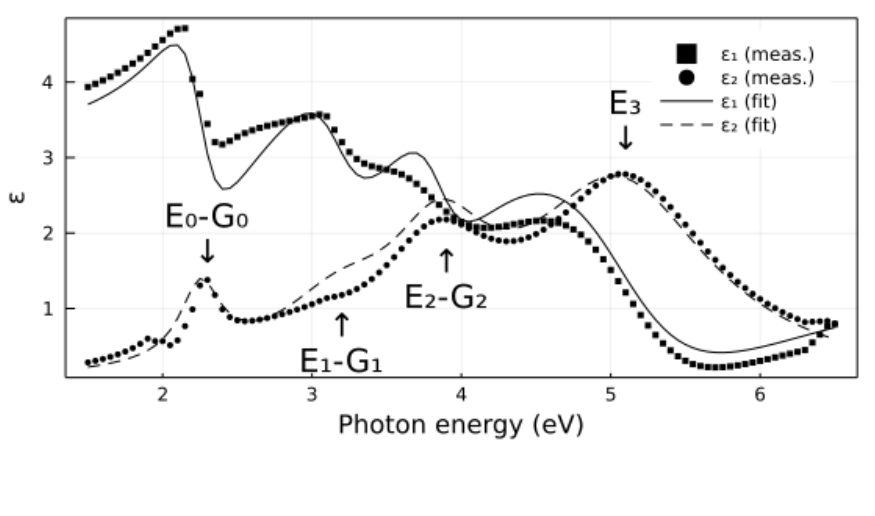}
\caption{Real (squares) and imaginary (circles) parts of the dielectric
functions of LNO thin film taken at RT. The symbols are the experimental
data, while the lines are calculated for real (solid) and imaginary
(dashed) parts using Eq. (1), Eq. (2), and $\epsilon_{1\infty} = 2$. }
\end{figure}

\section{4. Results and Discussion}\label{results-and-discussion}

In Fig. 2, we plot the real ($\epsilon$\textsubscript{1}) and imaginary
($\epsilon$\textsubscript{2}) parts of dielectric function spectra $\epsilon$($\omega$) of LNO
determined at RT. As seen in the figure, the experimental data reveal
clear structures at the 2.2 to 2.4-eV region. This structure originates
from transition at the \emph{E}\textsubscript{0} edge. The structures
appearing at the 3.0 to 3.4, 3.6 to 4.2, and 4.7 to 5.5-eV regions are due to
the \emph{E}\textsubscript{1}, \emph{E}\textsubscript{2}, and
\emph{E}\textsubscript{3} transitions, respectively.

The model dielectric function (MDF) approach given in Sec. IIB was used
to fit the experimental dispersion of $\epsilon$($\omega$) over the entire range of the
measurements (0 to 6.5 eV). The parameters such as
\emph{A}\textsubscript{0} and \emph{A}\textsubscript{1} are used as
adjustable constants for the calculations of $\epsilon$\textsubscript{1}($\omega$) and
$\epsilon$\textsubscript{2}($\omega$). As has been already mentioned in Sec. IIB, the
$\epsilon$\textsubscript{1}($\omega$) for LNO is, in overall, a monotonically decreasing
function with upward convex in 1.5 to 5-eV range except for several
optical anomalies such as \emph{E}\textsubscript{0}. This is very
peculiar compared to those for many other elemental and compound
semiconductors {[}20,22-25{]}. This is somehow reminiscent of that in
$\alpha$-Sn {[}24{]}. In this case, very strong absorption band is present in
the $\epsilon$\textsubscript{2} spectrum, giving rise to consistent explanation
in terms of the 3D \emph{M}\textsubscript{0} CP with very large
transition strength. This is not the case for LNO. To reproduce the
$\epsilon$\textsubscript{1} behavior, there seems no other solution except for
positive adoption of the \emph{M}\textsubscript{1}-type CP even for the
lowest energy gap because the difference between the $\epsilon$\textsubscript{1}
values at both ends is larger in the case of 3D
\emph{M}\textsubscript{1} (2D \emph{M}\textsubscript{0}). Comparison of
the behavior at both ends revealed us that the lineshape of the 2D
\emph{M}\textsubscript{0} MDF is known to drop significantly, while that
of the 3D \emph{M}\textsubscript{0} MDF remains relatively flat
{[}20{]}. Therefore, the 2D \emph{M}\textsubscript{0} MDF is more
suitable to reproduce the overall $\epsilon$\textsubscript{1} tendency. The
experimental data on $\epsilon$\textsubscript{1}($\omega$) turned out to be still
somehow larger than the model fit. To improve this fit, we then
introduced a phenomenological term $\epsilon_{1\infty}$ in addition to
$\epsilon$\textsubscript{1}($\omega$) to account for the contributions from higher-lying
energy gaps. This term $\epsilon_{1\infty}$ is assumed to be
nondispersive.

The solid and dashed lines in Fig. 2 are obtained from the sum of Eqs.
(1) and (2), and $\epsilon_{1\infty}$=2.0. The vertical arrows in the
figure indicate the excitonic peak energies
(\emph{E}\textsubscript{0}-\emph{G}\textsubscript{0},
\emph{E}\textsubscript{1}-\emph{G}\textsubscript{1},
\emph{E}\textsubscript{2}-\emph{G}\textsubscript{2}) and the positions
of the CP (\emph{E}\textsubscript{3}). The best-fit parameters are
listed in Table I. Due to the earlier-mentioned assumptions, we could
not determine the excitonic binding energies such as
\emph{G}\textsubscript{0} experimentally. Individual contributions to
the dielectric functions of the various energy gaps for LNO are shown in
Figs. 3 and 4, respectively. They are obtained from Eq. (1) for the
2D-exciton contribution in the \emph{E}\textsubscript{0},
\emph{E}\textsubscript{1}, and \emph{E}\textsubscript{2} regions, and
from Eq. (2) for the \emph{E}\textsubscript{3} gap contribution.

\begin{figure}
\includegraphics[width=0.6\textwidth]{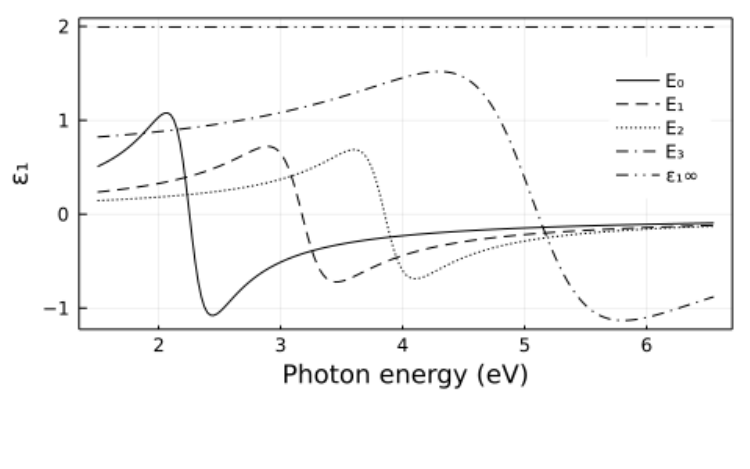}
\caption{ Individual contribution to imaginary part of dielectric function
of the various energy gaps for LNO. They are obtained from Eq. (1) for
the \emph{M}\textsubscript{1} (i.e., the 2D-exciton) contribution, and
from Eq. (2) for the \emph{E}\textsubscript{3}-gap contribution.}
\end{figure}

\begin{figure}
\includegraphics[width=0.6\textwidth]{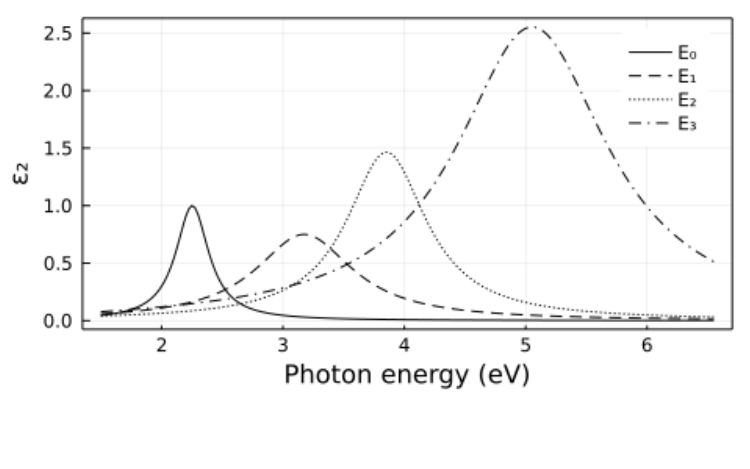}
\caption{ As of Fig. 3, but for real part. }
\end{figure}

The sums of Eqs. (1) and (2) fit the main features of the
$\epsilon$\textsubscript{1} and $\epsilon$\textsubscript{2} data, but the agreement with
the experiment is rather qualitative. The difference probably could come
from several sources, including (1) the Lorentz approximation for
broadening (which is known to give too much absorption below direct band
edges {[}31{]}), (2) the neglection for the excited-states excitons
{[}20{]}, or (3) the parabolic assumption of the CP's {[}19{]}.

As understood from the table, some of the best-fit parameters from the
$\epsilon$\textsubscript{1} analysis are different from the $\epsilon$\textsubscript{2}
counterparts. Being compliant with SCP theory's policy, we took the
average from these values, as shown in this figure. We believe that
these averaged values should be material parameters of LNO. An increase
in the number of MDF's (CP's) is expected to lead to an improved
agreement with the experiment, even with the same (common) values on the
parameters. However, it might be an excessive assumption to include an
additional transition to the position where no peak or anomaly is
observed. Indeed, this might be justified by the fact that calculated
SECI spectra include at least five CP's in the energy range (1.5 to 6.5
eV), as can be viewed from Fig. 5.

\begin{figure}
\includegraphics[width=0.6\textwidth]{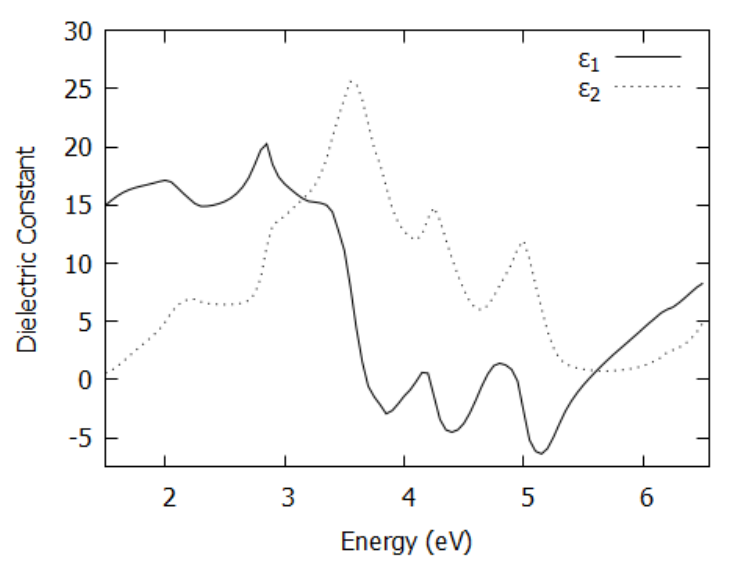}
\caption{Real and imaginary parts of the dielectric functions of LNO,
obtained from the \emph{ab-initio} calculation. }
\end{figure}

Neglecting contributions from the optical anomalies, overall
$\epsilon$\textsubscript{1}($\omega$) in the SECI spectrum is a monotonically decreasing
function with upward convex in the 1.5 to 4-eV range. This feature is in
good agreement with the experiment. Besides, the agreement on the
\emph{E}\textsubscript{0} energy is also good. According to the
\emph{ab-initio} calculation under the LDA level, the bandgap energy is
evaluated to be approximately 1.5 eV. The calculated energy is
significantly lower than that of experimental value (\emph{ca.} 2.3 eV).
As is often the case, it is well-known that the LDA level calculation
tends to underestimate the bandgap energy. On the other hand, the
\emph{ab-initio} SECI calculation considers the electron-hole
interaction (excitonic) effects in this material, leading to the
improved agreement with the experiment. The spectral blue-shift from the
LDA-level result is probably due to the so-called self-energy effect.
Furthermore, the energetic positions of \emph{E}\textsubscript{1} and
\emph{E}\textsubscript{3} CP's are in reasonable agreement with the
experiment.

We notice that the calculated spectra structures are sharper and more
distinct than those in the measured one. The calculated
$\epsilon$\textsubscript{1}($\omega$) amplitude as defined by the difference of the
$\epsilon$\textsubscript{1} value at 1.5 and 6.5 eV is also larger. These looser
agreements are probably related to the same origin. The broadening
parameter corresponding to the width of the Green functions used in the
calculation is significantly smaller than that obtained in the
measurement. The doublet splitting of the \emph{E}\textsubscript{2} is
observed in the SECI spectra. Experimentally, the double seems not
resolved due to the broadening effect. Thus, our results should
stimulate further work for this material both experimentally and
theoretically.

Finally, we mention refractive-index dispersion in the transparency
region. As seen in Fig. 2, the analysis results are in agreement with
the experiment ($\epsilon$\textsubscript{1}) over the entire range of photon
energies. On the other hand, the agreement is looser. An increase in the
parameter $\epsilon_{1\infty}$ from 2 to 3.5 improves the agreement in
the transparency region. However, this does not give a satisfactory fit
to the experimental data at higher photon energies (\textgreater{} 3.0
eV). Several research groups have reached a similar conclusion on GaP,
AlSb, and ZnSe. The looser agreement is probably related to the
earlier-mentioned Lorentz approximation for broadening {[}20,32-33{]}.
Indeed, Pollak's group has introduced a phenomenological linear cutoff
to eliminate this problem giving too much absorption below direct band
edges {[}31{]}. We did not attempt this here because the cutoff's
adoption forced us a numerical Kramers-Kronig conversion to the
$\epsilon$\textsubscript{1} counterpart, thus making it impossible to analyze the
experimental data within the regression analysis framework based on a
conventional Levenberg--Marquardt algorithm.

\begin{table}
\caption{Table I: Material parameters used in the calculation of the optical
constants of LNO.}
\begin{tabular}{cccc}
Parameters  &Values from $\epsilon$\textsubscript{1}  & Values from $\epsilon$\textsubscript{2}  & Average\\
\hline
\emph{E}\textsubscript{0} (eV) & 2.25 & 2.28 & 2.27 \\
$\Gamma$\textsubscript{0} (eV) & 0.19 & 0.17 & 0.18 \\
$A$\textsubscript{0} (eV) & 0.41 & 0.17 & 0.29 \\
\emph{E}\textsubscript{1} (eV) & 3.18 & 3.18 & 3.18 \\
$\Gamma$\textsubscript{1} (eV) & 0.28 & 0.49 & 0.39 \\
$A$\textsubscript{1} (eV) & 0.41 & 0.37 & 0.39 \\
\emph{E}\textsubscript{2} (eV) & 3.85 & 3.85 & 3.85 \\
$\Gamma$\textsubscript{2} (eV) & 0.25 & 0.40 & 0.33 \\
$A$\textsubscript{2} (eV) & 0.34 & 0.59 & 0.47 \\
\emph{E}\textsubscript{3} (eV) & 5.1 & 5.2 & 5.2 \\
$\Gamma$\textsubscript{3} & 0.3 & 0.3 & 0.3 \\
$A$\textsubscript{3} & 0.76 & 0.75 & 0.76 \\
$\epsilon_{1\infty}$ & 2 & & 2 \\
\hline
\end{tabular}
\end{table}

\section{5. Conclusion}\label{conclusion}

In summary, we have reported the dielectric functions of LNO around the
fundamental band gap \emph{E}\textsubscript{0} (\emph{ca.} 2.3 eV) and
up to 6.5 eV for RT. The observed spectra reveal distinct structures at
energies the \emph{E}\textsubscript{0}, \emph{E}\textsubscript{1},
\emph{E}\textsubscript{2} and \emph{E}\textsubscript{3} CP's. These data
are analyzed based on a simplified model of the interband transitions,
including these transitions as the main dispersion mechanisms. The
analyzed results are in qualitative agreement with the experiment. Many
important material parameters of these transitions were determined. The
comparison was made with the results of \emph{ab-initio} calculations,
taking the electron-hole correlation effects into account.

\section{Acknowledgments}\label{acknowledgments}

T.M. wishes to thank the financial support of the Ministry of Education,
Culture, Sports, Science and Technology Grant No. KAKENHI-19K05303. This
work was partially supported from No.~KAKENHI-20K15169. All of the
\emph{ab-initio} calculations were performed in the Supercomputer Center
at the Institute of Solid-State Physics, the University of Tokyo.

\section{References}\label{references}

{[}1{]} G. Meyer, R. Hoppe, The first oxoniobate LiNbO\textsubscript{2}, Angew. Chem.
Int. Ed. 13 (1974) 744--745.
\url{https://doi.org/10.1002/anie.197407441}.

{[}2{]} N. Kumada, S. Muramatu, F. Muto, N. Kinomura, S. Kikkawa, M.
Koizumi, Topochemical reactions of LiNbO\textsubscript{2}, J. Solid State Chem. 73
(1988) 33--39. \url{https://doi.org/10.1016/0022-4596(88)90050-3}.

{[}3{]} N. Sarmadian, R. Saniz, B. Partoens, D. Lamoen, Easily doped
p-type, low hole effective mass, transparent oxides, Sci. Reports. 6
(2016) 20446. \url{https://doi.org/10.1038/srep20446}.

{[}4{]} W.E. Henderson, W.L. Calley, A.G. Carver, H. Chen, W.A.
Doolittle, A versatile metal-halide vapor chemistry for the epitaxial
growth of metallic, insulating and semiconducting films, J. Cryst.
Growth. 324 (2011) 134--141.
\url{https://doi.org/10.1016/j.jcrysgro.2011.03.049}.

{[}5{]} M.B. Tellekamp, J.C. Shank, W.A. Doolittle, Molecular beam
epitaxy of lithium niobium oxide multifunctional materials, J. Cryst.
Growth. 463 (2017) 156.
\url{https://doi.org/10.1016/j.jcrysgro.2017.02.020}.

{[}6{]} T. Soma, K. Yoshimatsu, A. Ohtomo, P-type transparent
superconductivity in a layered oxide, Science Adv. 6 (2020) eabb8570.
\url{https://doi.org/10.1126/sciadv.abb8570}.

{[}7{]} M.J. Geselbracht, T.J. Richardson, A.M. Stacy, Superconductivity
in the layered compound LiNbO\textsubscript{2}, Nature (London). 345 (1990) 324.
\url{https://doi.org/10.1038/345324a0}.

{[}8{]} M.A. Rzeznik, M.J. Geselbracht, M.S. Thompson, A.M. Stacy,
Superconductivity and phase separation in NaNbO\textsubscript{2}, Angew. Chem. Int. Ed.
32 (1993) 254--255. \url{https://doi.org/10.1002/anie.199302541}.

{[}9{]} E.G. Moshopoulou, P. Bordet, J.J. Capponi, Superstructure and
superconductivity in LiNbO\textsubscript{2} single crystals, Phys. Rev. B. 59 (1999)
9590--9599. \url{https://doi.org/10.1103/physrevb.59.9590}.

{[}10{]} G.T. Liu, J.L. Luo, Z. Li, Y.Q. Guo, N.L. Wang, D. Jin, T.
Xiang, Evidence of s-wave pairing symmetry in the layered superconductor
LiNbO\textsubscript{2} from specific heat measurements, Phys. Rev. B. 74 (2006).
\url{https://doi.org/10.1103/physrevb.74.012504}.

{[}11{]} J.D. Greenlee, C.F. Petersburg, W.L. Calley, C. Jaye, D.A.
Fischer, F.M. Alamgir, W.A. Doolittle, In-situ oxygen x-ray absorption
spectroscopy investigation of the resistance modulation mechanism in
LiNbO\textsubscript{2} memristors, Appl. Phys. Lett. 100 (2012) 182106.
\url{https://doi.org/10.1063/1.4709422}.

{[}12{]} S.A. Howard, C.N. Singh, G.J. Paez, M.J. Wahila, L.W. Wangoh,
S. Sallis, K. Tirpak, Y. Liang, D. Prendergast, M. Zuba, J. Rana, A.
Weidenbach, T.M. McCrone, W. Yang, T.-L. Lee, F. Rodolakis, W.
Doolittle, W.-C. Lee, L.F.J. Piper, Direct observation of delithiation
as the origin of analog memristance in LiNbO\textsubscript{2}, APL Materials. 7 (2019)
071103. \url{https://doi.org/10.1063/1.5108525}.

{[}13{]} X. Xu, G. Liu, S. Ni, J.T.S. Irvine, Layered lithium niobium
oxide LiNbO\textsubscript{2} as a visible-light-driven photocatalyst for H\textsubscript{2} evolution,
J. Phys.: Energy. 1 (2018) 015001.
\url{https://doi.org/10.1088/2515-7655/aad4be}.

{[}14{]} E.R. Ylvisaker, W.E. Pickett, First-principles study of the
electronic and vibrational properties of LiNbO\textsubscript{2}, Phys. Rev. B. 74 (2006)
075104. \url{https://doi.org/10.1103/physrevb.74.075104}.

{[}15{]} D.L. Novikov, V.A. Gubanov, V.G. Zubkov, A.J. Freeman,
Electronic structure and electron-phonon interactions in layered Li\textsubscript{x}NbO\textsubscript{2}
and Na\textsubscript{x}NbO\textsubscript{2}, Phys. Rev. B. 49 (1994) 15830.
\url{https://doi.org/10.1103/physrevb.49.15830}.

{[}16{]} K.-W. Lee, J. Kunes, R.T. Scalettar, W.E. Pickett, Correlation
effects in the triangular lattice single-band system Li\textsubscript{x}NbO\textsubscript{2}, Phys. Rev.
B. 76 (2007) 144513. \url{https://doi.org/10.1103/physrevb.76.144513}.

{[}17{]} J.U. Rahman, N.V. Du, G. Rahman, V.M. Garcia-Suarez, W.-S. Seo,
M.H. Kim, S. Lee, Localized double phonon scattering and DOS induced
thermoelectric enhancement of degenerate nonstoichiometric LiNbO\textsubscript{2}
compounds, RSC Advances. 7 (2017) 53255.
\url{https://doi.org/10.1039/c7ra10557f}.

{[}18{]} M.J. Geselbracht, A.M. Stacy, A.R. Garcia, B.G. Silbernagel,
G.H. Kwei, Local environment and lithium ion mobility in lithium niobate
LiNbO\textsubscript{2}: Inferences from structure, physical properties, and NMR, J.
Phys. Chem. 97 (1993) 7102. \url{https://doi.org/10.1021/j100129a030}.

{[}19{]} S. Loughin, R. French, L. DeNoyer, W.-Y. Ching, Y.-N. Xu,
Critical point analysis of the interband transition strength of
electrons, J. Phys. D. 29 (1996) 1740.
\url{https://doi.org/10.1088/0022-3727/29/7/009}.

{[}20{]} S. Adachi, T. Taguchi, Optical properties of ZnSe, Phys. Rev.
B. 43 (1991) 9569. \url{https://doi.org/10.1103/PhysRevB.43.9569}.

{[}21{]} Y. Petroff, M. Balkanski, Coulomb effects at saddle-type
critical points in CdTe, ZnTe, ZnSe, and HgTe, Phys. Rev. B. 3 (1971)
3299. \url{https://doi.org/10.1103/physrevb.3.3299}.

{[}22{]} S. Adachi, Model dielectric constants of GaP, GaAs, GaSb, InP,
InAs, and InSb, Phys. Rev. B. 35 (1987) 7454.
\url{https://doi.org/10.1103/PhysRevB.35.7454}.

{[}23{]} S. Adachi, Model dielectric constants of si and ge, Phys. Rev.
B. 38 (1988) 12966. \url{https://doi.org/10.1103/PhysRevB.38.12966}.

{[}24{]} S. Adachi, Optical properties of alpha-sn, J. Appl. Phys. 66
(1989) 813. \url{https://doi.org/10.1063/1.343502}.

{[}25{]} S. Ninomiya, S. Adachi, Optical properties of cubic and
hexagonal CdSe, J. Appl. Phys. 78 (1995) 4681.
\url{https://doi.org/10.1063/1.359815}.

{[}26{]} R.E. Denton, R.D. Campbell, S.G. Tomlin, The determination of
the optical constants of thin films from measurements of reflectance and
transmittance at normal incidence, J. Phys. D. 5 (1972) 852.
\url{https://doi.org/10.1088/0022-3727/5/4/329}.

{[}27{]} P. Giannozzi, S. Baroni, N. Bonini, M. Calandra, R. Car, C.
Cavazzoni, D. Ceresoli, G.L. Chiarotti, M. Cococcioni, I. Dabo, A.D.
Corso, S. de Gironcoli, S. Fabris, G. Fratesi, R. Gebauer, U. Gerstmann,
C. Gougoussis, A. Kokalj, M. Lazzeri, L. Martin-Samos, N. Marzari, F.
Mauri, R. Mazzarello, S. Paolini, A. Pasquarello, L. Paulatto, C.
Sbraccia, S. Scandolo, G. Sclauzero, A.P. Seitsonen, A. Smogunov, P.
Umari, R.M. Wentzcovitch, QUANTUM ESPRESSO: A modular and open-source
software project for quantum simulations of materials, J. Phys.: Cond.
Mat. 21 (2009) 395502.
\url{https://doi.org/10.1088/0953-8984/21/39/395502}.

{[}28{]} P. Giannozzi, O. Andreussi, T. Brumme, O. Bunau, M.B. Nardelli,
M. Calandra, R. Car, C. Cavazzoni, D. Ceresoli, M. Cococcioni, N.
Colonna, I. Carnimeo, A.D. Corso, S. de Gironcoli, P. Delugas, R.A.D.
Jr, A. Ferretti, A. Floris, G. Fratesi, G. Fugallo, R. Gebauer, U.
Gerstmann, F. Giustino, T. Gorni, J. Jia, M. Kawamura, H.-Y. Ko, A.
Kokalj, E. Kucukbenli, M. Lazzeri, M. Marsili, N. Marzari, F. Mauri,
N.L. Nguyen, H.-V. Nguyen, A. Otero-de-la-Roza, L. Paulatto, S. Ponce,
D. Rocca, R. Sabatini, B. Santra, M. Schlipf, A.P. Seitsonen, A.
Smogunov, I. Timrov, T. Thonhauser, P. Umari, N. Vast, X. Wu, S. Baroni,
Advanced capabilities for materials modelling with QUANTUM ESPRESSO, J.
Phys.: Cond. Mat. 29 (2017) 465901.
\url{https://doi.org/10.1088/1361-648X/aa8f79}.

{[}29{]} K. Nakamura, Y. Yoshimoto, R. Arita, S. Tsuneyuki, M. Imada,
Optical absorption study by ab initio downfolding approach: Application
to GaAs, Phys. Rev. B. 77 (2008) 195126.
\url{https://doi.org/10.1103/physrevb.77.195126}.

{[}30{]} K. Nakamura, Y. Yoshimoto, Y. Nomura, T. Tadano, M. Kawamura,
T. Kosugi, K. Yoshimi, T. Misawa, Y. Motoyama, RESPACK: An ab initio
tool for derivation of effective low-energy model of material,
arXiv:2001.02351. (2020). \url{https://arxiv.org/abs/2001.02351}.

{[}31{]} T. Holden, P. Ram, F.H. Pollak, J.L. Freeouf, B.X. Yang, M.C.
Tamargo, Spectral ellipsometry investigation of ZnCdSe lattice matched
to InP, Phys. Rev. B. 56 (1997) 4037--4046.
\url{https://doi.org/10.1103/PhysRevB.56.4037}.

{[}32{]} K. Strössner, S. Ves, M. Cardona, Refractive index of GaP and
its pressure dependence, Phys. Rev. B. 32 (1985) 6614.
\url{https://doi.org/10.1103/physrevb.32.6614}.

{[}33{]} S. Zollner, C. Lin, E. Schönherr, A. Böhringer, M. Cardona, The
dielectric function of AlSb from 1.4 to 5.8 eV determined by
spectroscopic ellipsometry, J. Appl. Phys. 66 (1989) 383.
\url{https://doi.org/10.1063/1.343888}.
\end{document}